\documentclass{article}

\usepackage{spconf}
\usepackage{microtype}
\usepackage{graphicx}
\usepackage{subfigure}
\usepackage{booktabs} 
\usepackage{amsmath}
\usepackage{multirow}
\usepackage{amssymb}
\usepackage{makecell}

\title{Voice and Accompaniment Separation in Music Using Self-attention Convolutional Neural Networks}

\name{Yuzhou Liu\textsuperscript{1}, Balaji Thoshkahna\textsuperscript{2}, Ali Milani\textsuperscript{3}, Trausti Kristjansson\textsuperscript{3}}
\address{\textsuperscript{1}Ohio State University; \textsuperscript{2}Amazon Music, Bangalore; \textsuperscript{3}Amazon Lab126, CA}

\begin{document}


\maketitle




\begin{abstract}

Music source separation has been a popular topic in signal processing for decades, not only because of its technical difficulty, but also due to its importance to many commercial applications, such as automatic karoake and remixing. 
In this work, we propose a  novel self-attention network to separate voice and accompaniment in music.
First, a convolutional neural network (CNN) with densely-connected CNN blocks is built as our base network.
We then insert self-attention subnets at different levels of the base CNN to make use of the long-term intra-dependency of music, i.e., repetition.
Within self-attention subnets, repetitions of the same musical patterns inform reconstruction of other repetitions, for better source separation performance. Results show the proposed method leads to 19.5\% relative improvement in vocals separation in terms of SDR. We compare our methods with state-of-the-art systems i.e. MMDenseNet \cite{MM} and MMDenseLSTM.\cite{MMLSTM}.

\end{abstract}


\section{Introduction}
\label{sec:1}

Music source separation is a fundamental problem in music information retrieval.
As an important branch of music source separation, voice and accompaniment separation in music has grabbed great attention recently.
It plays a crucial role in karaoke creation, music remix and related commercial applications, especially when the clean music sources are not available, e.g., live recording.

With the rapid development of deep learning, more and more researchers start to formulate voice and accompaniment separation as a regression problem. 
The general idea is to feed spectral features into neural networks (NNs) for source estimation.
Some of the earlier NN based studies \cite{VADNN1, VADNN2, VADNN3} adopt feedforward fully connected neural networks (FNNs) or long short-term memory (LSTM) recurrent neural networks (RNNs) for this task. 
They have shown tremendous improvement over conventional model-based approaches like non-negative matrix factorization (NMF).
Convolutional neural networks (CNNs) have also been explored recently, where a time-frequency (T-F) representation of music is fed as 2-D single-channel input.
In \cite{UNET1, UNET2}, a UNet-CNN structure is utilized, which consists of a sequence of downsampling convolutional layers for encoding, and upsampling convolutional layers for decoding. Skip connections are added between layers at the same level in the encoder and decoder to preserve raw information. 
In \cite{MM}, each convolutional layer in the UNet-CNN is replaced by a densely-connected CNN block, which consists of a series of convolutional layers, with each layer connected to every other layer in a feed-forward fashion.
This Dense-UNet structure alleviates the vanishing-gradient problem, encourages feature reuse, and substantially reduces the number of parameters in UNet-CNNs.
In \cite{MMLSTM}, BLSTM RNNs are further added to Dense-UNet to incorporate more temporal context, denoted by MMDenseLSTM.

In the remainder of this paper, the proposed method is described in Section 2.
In Section~\ref{sec:3}, we present experimental results and comparisons.
A conclusion is given in Section~\ref{sec:4}.

\section{System description}
\label{sec:2}
In this section, we first introduce Dense-UNet as our base network architecture.
Then, we propose self-attention subnets, and insert them at different levels of the base network. 
More details about the complete architecture are presented in the end.

\subsection{Dense-UNet for voice and accompaniment separation}
\label{sec:2.1}

The goal of Dense-UNet is is to predict the voice and accompaniment component in a music mixture based on masking: the output of the final layer is a soft mask that is multiplied element-wise with the magnitude STFT of the mixture to obtain the final estimate. The detailed architecture of Dense-UNet in given in Fig. 1. We first describe its input and output as follows.

Let \(X_1(t,f)\) denote the short-time discrete Fourier transform (STFT) of human voice, where \(t\) and \(f\) are the time and frequency indices. Let \(X_2(t,f)\) denote the STFT of accompaniment. The music mixture can be defined as:
\begin{equation}
Y(t,f)=X_1(t,f)+X_2(t,f)
\end{equation}

We feed the magnitude STFT of the mixture signal \(|Y(t,f)|\) into a Dense-UNet to predict two masks \(M_1(t,f)\) and \(M_2(t,f)\) for the two sources.
The masks are then multiplied with the mixture signal to reconstruct the original sources:
\begin{equation}
|\tilde{X_i}(t,f)|=M_i(t,f)\odot|Y(t,f)|, \quad i = 1, 2
\end{equation}
Here \(\odot\) denotes element-wise multiplication, and \(|\tilde{X_i}(t,f)|\) is the reconstructed magnitude STFT of a source.
\(|\tilde{X_i}(t,f)|\) is then coupled with mixture phase to resynthesize the time-domain signal of a source. 

The loss function used to train the model is the \(l_1\) norm of the difference of the target magnitude STFT and the masked magnitude STFT:
\begin{equation}
J_t=\sum_{i}\sum_{t,f}(\lVert M_i(t,f)\odot|Y(t,f)|-|X_i(t,f)| \lVert 
\end{equation}
where \(\lVert \cdot \lVert\) denotes \(l_1\) norm.

\begin{figure}[]
\begin{minipage}[a]{\linewidth}
  \centering
  \centerline{\includegraphics[width=0.95\linewidth]{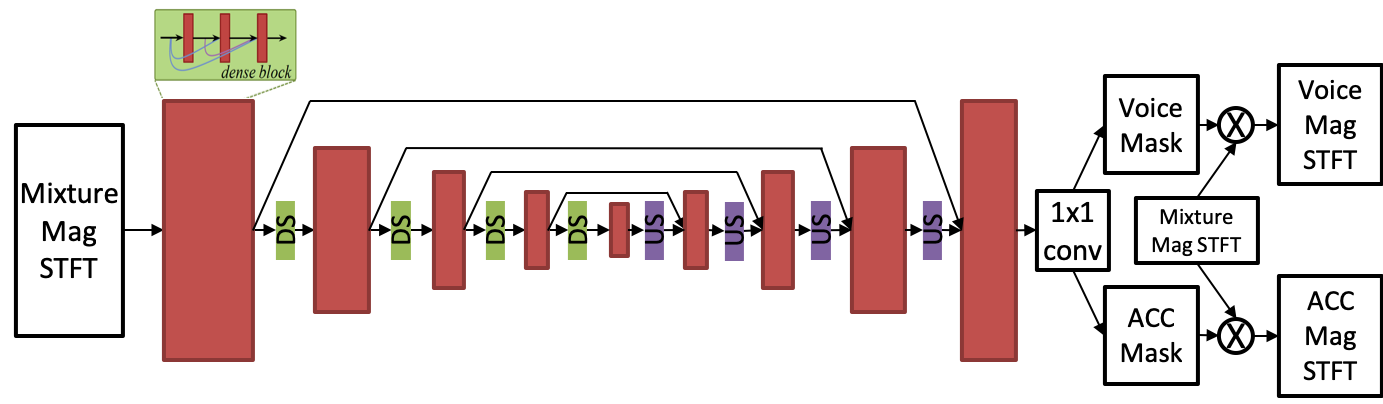}}
\medskip
\end{minipage}
\caption{Diagram of the Dense-UNet. Red blocks denote dense CNN blocks. Green blocks denote max pooling layers. Purple blocks denotes transpose convolutional layers. Skip connections are added to connect layers at the same level.}
  \vspace{-0.9em}
\label{fig:1}
\end{figure}

\begin{figure*}[]
\begin{minipage}[a]{\linewidth}
  \centering
  \centerline{\includegraphics[width=0.7\linewidth]{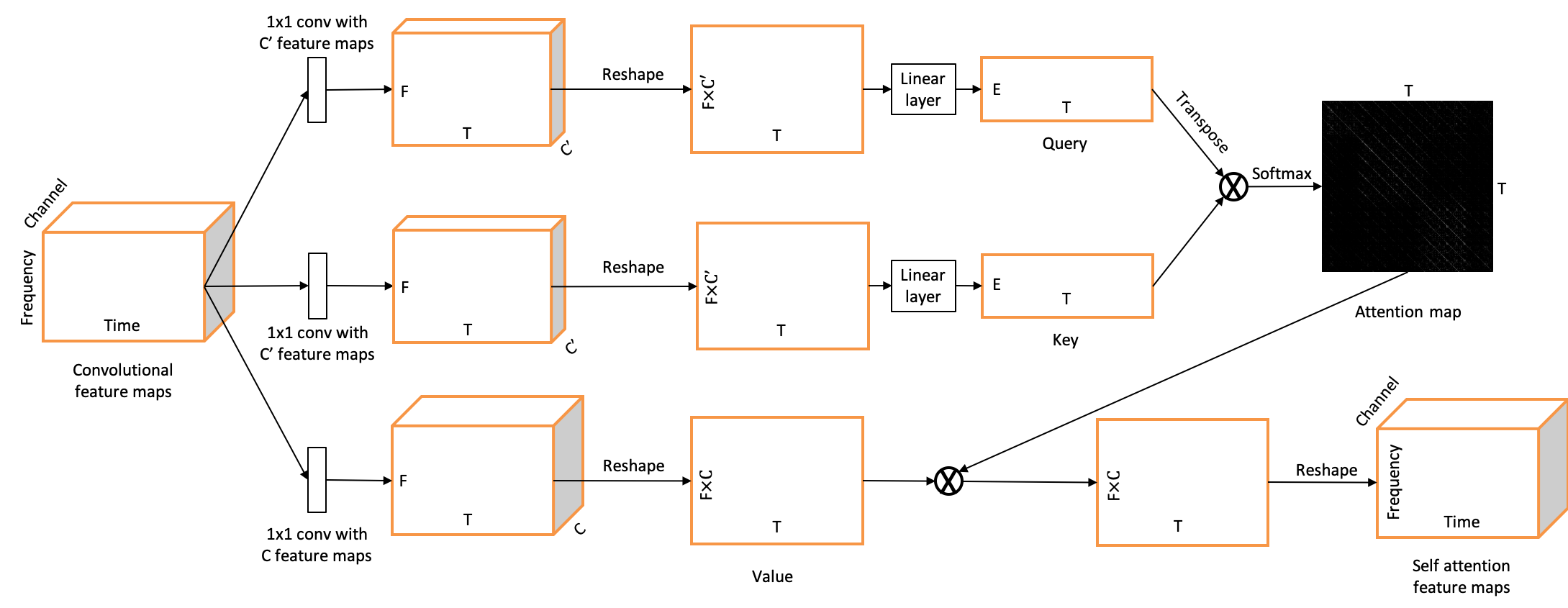}}
\medskip
\end{minipage}
\caption{The proposed self attention mechanism. The 
 \(\otimes\) denotes matrix multiplication. The softmax
operation is performed on each row. The linear layer transform feature dimension from \(F\times C'\) per time segment to \(E\). The dimensions of different feature maps are marked on the figure.}
  \vspace{-0.9em}
\label{fig:1}
\end{figure*}

\begin{figure}[]
\begin{minipage}[a]{\linewidth}
  \centering
  \centerline{\includegraphics[width=0.95\linewidth]{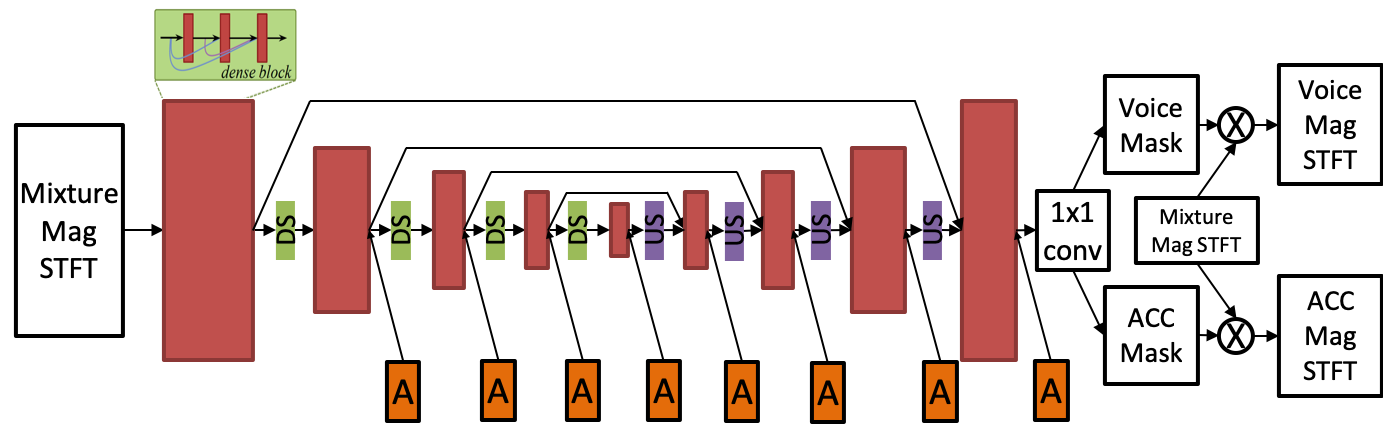}}
\medskip
\end{minipage}
\caption{Diagram of the self-attention Dense-UNet. Orange blocks marked ‘A’ denote self-attention subnets.}
  \vspace{-0.9em}
\label{fig:1}
\end{figure}

%

The Dense-UNet we implemented, as shown in Figure 1,  is based on an UNet architecture \cite{UNET2}.
It consists a series of convolutional layers, downsampling layers and upsampling layers.
The first half of the network corresponds to the encoding process, which encodes input features into a higher level of abstraction.
In this half, we alternate convolutional layers and downsampling layers, allowing the network to model longer contexts and wider frequency range dependency at a low computational cost.
In the second half of the network, namely the decoding network, we alternate convolutional layers and upsampling layers to project the encoded feature maps back to its original resolution.
In this paper,  we use a strided \(2\times 2\) max pooling layer as the downsampling layer. Strided transpose convolutional layers are used as upsampling layers. 
The number of channels stays unchanged after both downsampling and upsampling layers.
To make the decoding network preserve a high level of details of the input, we add skip connections between layers at the same hierarchical level in the encoder and decoder.

In the next step, we replace convolutional layers in the original UNet with densely-connected CNN blocks (DenseNet) \cite{DENSE}. 
The basic idea of DenseNet is to decomposes one convolutional layer with many channels into a sequence of densely connected convolutional layers with less channels, where each layer is connected to every other layer in a feed-forward fashion:
\begin{equation}
x_l = H_{l} ([x_{l-1},x_{l-2},...,x_0])
\end{equation}
where \(x_0\) denotes the input feature map, \(x_l\) denotes the output of the \(l^{th}\) layer, \([...]\) denotes concatenation, \(H_l\) denotes the \(l^{th}\) convolutional layer followed by ELU activation. 
The DenseNet structure enables all intermediate layers to directly receive gradient from the output. It also encourages feature reuse, thus reduces the total number of parameters.
It has shown excellent performance in image classification and music source separation.
In this study, all output layers \(x_l\) in a dense block has the same number of channels as the input, denoted by C. 
And the total number of layers in each dense block is denoted by K.
As shown in Figure 1, we alternate 9 dense blocks with 4 downsampling layers and 4 upsampling layers.
After the last dense block, we use a \(1 \times 1 \) convolution to reorganize the feature map, and then output the voice mask and accompaniment mask for STFT domain source reconstruction.

It should be noted that our base Dense-UNet model is a simplified version of MMDenseLSTM.
In MMDenseLSTM, the authors split frequency into multiple bands, and build separate models for different bands.
They also insert BLSTM RNNs into Dense-UNet for a longer temporal context. 

\subsection{Self-attention Dense-UNet}
\label{sec:2.2}

Musical pieces are usually composed of structures where a singer overlays varying lyrics on a repeating accompaniment.
Therefore, for one specific time segment, if we can find other time segments sharing the same accompaniment or rhythm, and take them as reference, we can definitely have a better separation for both voice and accompaniment. 
The REPET algorithm \cite{REPET} was built upon this idea.
However, rhythm and repetition of accompaniment usually have very long-range dependencies in the time domain.
For example, drum beats might have repetitive patterns every couple seconds. The same chord of guitar might be played every 10 seconds.
These dependencies are too long for Dense-UNet to handle due to its local CNN filters. 
Even enhanced with BLSTM RNNs, MMDenseLSTM still can not handle skipped repetitions since it only focuses on smooth local memory.

In this paper, we introduce the self attention mechanism \cite{ST1, ST2} to model the repetition of music. 
Self attention calculates the response at a position in a sequence by attending to all positions within the same sequence.
It has been successfully applied in fields like machine translation and image generation \cite{ST1, ST2}.
In this study, we use self attention to relate each time segment with other time segments sharing the same repetitive patterns, and use these repetitive patterns as additional information for source separation.

Our self-attention subnet is designed as in Figure 2. Convolutional output of Dense-UNet at different levels is fed as input to self-attention subnets. The input feature map has three dimensions: channel, frequency and time, denoted by \(C\times F\times T\), where T covers a long time range (around 20 seconds in this study), F covers the whole frequency range of the spectrogram, C corresponds to the the number of channels in the output of a dense block.
It should be noted that the resolution of \(T\) varies at different levels of Dense-UNet, so we refer the time unit in the feature map as time segment instead of time frame.

We first feed the input feature map into three \(1\times 1\) convolutional layers.
The first two \(1\times 1\)  conv layers reduce the number of channels to \(C'\), and they transform the input into a feature space to compute attention, or similarity between time segments.
The last \(1\times 1\) conv layer keeps the number of channels in the input. It transforms the input to the output feature space.

Since we want to compute attention or similarity only between time segments, we reshape all three feature maps to 2-D representations, with the second dimension as time. The first two feature maps are then linearly transformed to an \(E\times T\) matrix for a higher level of abstraction and dimensionality reduction. They are denoted by Query and Key respectively. We can use \(\bold{Q}(t) \in \mathbb{R}^{E\times 1} \) and \(\mathbf{K}(t) \in \mathbb{R}^{E \times 1}\) to represent each time segment of Query and Key.
The third feature map is reshaped into a 2-D matrix, denoted by Value or \(\bold{V} \in \mathbb{R}^{(C\times F)\times T} \).
A time segment of \(\bold{V}\) is denoted by \(\bold{V}(t) \in \mathbb{R}^{(C\times F)\times 1} \).

The attention map is calculated as:
\begin{equation}
\beta_{i,j} = \frac{exp(s_{ij})}{\sum_{j=1}^{T}{exp(s_{ij})}}, \text{where     } s_{ij} = \bold{Q}(i)^T\bold{K}(j)
\end{equation}
\(\beta_{i,j}\) indicates how much information the \(j^{th}\) time segment can contribute when synthesizing the  \(i^{th}\) time segment.
In other words, it measures the similarity between time segments implicitly.

The output of the attention subnet is 
\begin{equation}
\bold{O} = (\bold{O}(1),\bold{O}(2),...,\bold{O}(T)) \in \mathbb{R}^{(C\times F)\times T}
\end{equation}
where
\begin{equation}
\bold{O}(i) = \sum_{j=1}^T{\beta_{i,j}\bold{V}(j)}
\end{equation}
In other words, the output \(\bold{O}\) is a weighted sum of the Value matrix, and the attention map serves as the weights. If a remote time segment is more related with the current segment, it has more weight on the current time segment in the output.
In the end, \(\bold{O}\) is reshaped to the 3-D representation of \(C\times F\times T\), concatenated with the input feature map, and fed to the next level.

As shown in Figure 3, we insert such self-attention subnets into different levels of Dense-UNet,  so that they can capture dependencies at different levels of abstraction. 
The insertion takes place after dense blocks and before downsampling/upsampling layers. We do not add any self-attention subnets for first dense block because the feature space is relatively raw at that level, and we want to keep those raw features for better reconstruction.


We should also point out that for some music genres where there is almost no repetition and intra-dependency, like jazz, self-attention still works.  
In this situation, the network will learn to pay attention only to the current frame, so that the attention map becomes close to an identity matrix.
For other music genres with strong repetitive patterns, the self-attention subnets work very well with little increase in computational cost.

\subsection{Architecture details}
\label{sec:2.3}

The architecture of Dense-UNet and self attention Dense-UNet is illustrated in Figure 1, 2 and 3.
Given the heavy computational requirements of training, we downsample all input audios to 16 kHz in order to speed up processing.
We then compute the STFT with a window size of 1024 and hop size of 256.
Magnitude STFT is used in both input and output, with no normalization.
For Dense-UNet, the valid input/output window size is set to 128 frames, more input is padded based on the padding style of the network.
For self-attention Dense-UNet, since we need a much longer context, 1250 frames is used as the valid input/output window size.
During test, we divide a whole song into multiple input windows with 3/4 overlap, and average results from different input windows. 

There are two output layers in the Dense-UNet, one for each source.
The ReLU activation \cite{ReLU} is used for mask estimation.
A \(1 \times 1\) convolutional layer is applied before the output layer for feature reorganization.
For the rest of Dense-UNet, we alternate 9 dense blocks with 4 downsampling layers and 4 upsampling layers. 
In each dense block, the number of channels C is set to 32, and the total number of dense layers K is set to 4.
Each dense layer has a kernel size of 3 and stride of 1. ELU is used as activation without any further normalization.
 The last layer in each dense block use valid padding, all other layers use same padding. Based this, we add temporal context and zero-pad along the frequency axis for the input.
 In self-attention subnets, the number of channels C' is set to 5, and the encoding dimension E for Query and Key is set to 20.

Both networks are trained using the ADAM \cite{Adam} optimizer with a learning rate of 5e-5. 

\section{Evaluation results and comparisons}
\label{sec:3}

To evaluate the proposed methods, we use the training partition of the DSD100 database \cite{DSD100} for training data generation. 
We randomly scale, shift and add sources from different songs for data augmentation.
In the end, 450 songs are generated for training, 400 from remix, and 50 from original database.
Next, we split the MedleyDB \cite{MedleyDB} and CCMixter \cite{CCMixter} database into thirds, and use one third of tracks from each database for cross validation. 
Our test set is built by taking another third of tracks from MedleyDB and CCMixter, in addition to 25 tracks from the test partition of DSD100. 
All songs are downsampled to 16 kHz in training, test and evaluation.
We calculate the track-wise (normalised) SDR, SIR, and SAR as metrics.
The average results on the test set are reported in Table I.

\begin{table}[!t]
\renewcommand{\arraystretch}{1.3}
\caption{Average test results on the proposed test set. A and V denote accompaniment and voice respectively. }
\label{TABLE3}
\centering
\resizebox{\columnwidth}{!}{
\begin{tabular}{c|c|c|c|c}
\Xhline{2.5\arrayrulewidth}
\multirow{2}{*}{Metric (dB)} & \multirow{2}{*}{Dense-UNet }& Self-attention &Absolute&Relative\\
 & &Dense-UNet&difference&difference\\
\hline
\hline

SDR (A)& 12.91 &14.10&1.19&9.22\%\\ 
\hline
SDR (V)  & 6.58 & 8.08&1.5&22.80\%\\ 
\hline
SIR(A)  & 17.28& 18.42&1.14&6.60\%\\ 
\hline
SIR(V)  &14.24 &15.44&1.2&8.43\%\\ 
\hline
SAR (A)& 15.31 & 16.50&1.19&7.77\%\\ 
\hline
SAR (V)& 7.94 & 9.34&1.4&17.63\%\\ 

\Xhline{2.5\arrayrulewidth}
\end{tabular}
}
\end{table}

As shown in the table, self-attention Dense-UNet sigificantly outperforms Dense-UNet in terms of almost all metrics. The gap between self-attention Dense-UNet and oracle magnitude STFT is only around 2-3 dB for all metrics.

We then compare our trained models with two state-of-the-art systems, i.e., MMDenseNet \cite{MM} and MMDenseLSTM \cite{MMLSTM}. The MMDenseNet is basically an extended version of our Dense-UNet with multi-band processing. In MMDenseLSTM, the authors further add BLSTM layers to incorporate longer temporal context. We directly test our trained models under the same test setup as in \cite{MMLSTM}. The comparison is given below. 

As shown in the table, MMDenseLSTM leads to relatively little improvement compared with MMDenseNet, reflecting the ineffectiveness of RNNs for this task. Our baseline model Dense-UNet outperforms MMDenseLSTM, probably because of different structures and training recipes. The proposed self-attention Dense-UNet improves our baseline by a large margin for both accompaniment and voice.

\begin{table}[!t]
\renewcommand{\arraystretch}{1.3}
\caption{Median test results on a public test set. A and V denote accompaniment and voice respectively. }
\label{TABLE5}
\centering
\resizebox{0.8\columnwidth}{!}{
\begin{tabular}{c|c|c}
\Xhline{2.5\arrayrulewidth}
Metric (dB) &SDR (A)&SDR (V)\\
\hline
\hline
MMDenseNet & 12.10&6.00\\ 
\hline
MMDenseLSTM  & 12.73 & 6.31\\
\hline
Dense-UNet  & 12.75& 6.46\\ 
\hline
Self-attention Dense-UNet &13.90 &7.72\\ 

\Xhline{2.5\arrayrulewidth}
\end{tabular}
}
\end{table}

\section{Conclusion}
\label{sec:4}
The self-attention mechanism improves SDR performance for vocals by more than 19.5\%  (1.17 dB absolute) and 9\% for accompaniment (1.26 dB absolute) over our non-attention system. Compared to the state of the art MMDenseLSTM and MMDenseLSTM, our best system improves accompaniment SDR by 14.9\% and 9.2\% respectively and voice SDR by 28.7\% and 22.3\% respectively.


\bibliographystyle{IEEEbib}
\bibliography{refs}

\end{document}